# Modelling plasticity of unsaturated soils in a thermodynamically consistent framework

Olivier COUSSY, Jean-Michel PEREIRA

*Université Paris-Est, UR Navier, École des Ponts ParisTech, Marne-la-Vallée, France.*

**ABSTRACT**

Constitutive equations of unsaturated soils are often derived in a thermodynamically consistent framework through the use a unique 'effective' interstitial pressure. This later is naturally chosen as the space averaged interstitial pressure. However, experimental observations have revealed that two stress state variables were needed to describe the stress-strain-strength behaviour of unsaturated soils. The thermodynamics analysis presented here shows that the most general approach to the behaviour of unsaturated soils actually requires three stress state variables: the suction, which is required to describe the retention properties of the soil and two effective stresses, which are required to describe the soil deformation at water saturation held constant. Actually, it is shown that a simple assumption related to internal deformation leads to the need of a unique effective stress to formulate the stress-strain constitutive equation describing the soil deformation. An elastoplastic framework is then presented and it is finally shown that the Barcelona Basic Model, a commonly accepted model for unsaturated soils, as well as all models deriving from it, appear as special extreme cases of the thermodynamic framework proposed here.

**INTRODUCTION**

Within the area of saturated soil mechanics, a pioneering work applying the mathematical theory of plasticity to soil mechanics may be attributed to Roscoe and his co-workers on the general concept of critical state [1]. This work resulted in the well-known Cam-clay model for saturated soils involving Terzaghi's effective stress. An extension of this model to unsaturated soils has been proposed by Alonso *et al.* [2] within a simple elastoplastic framework. This extension points out the need of two independent state variables to capture the experimentally observed behaviour of unsaturated soils instead of a unique stress variable (effective stress) in the case of saturated states.

Université Paris-Est, UR Navier, École des Ponts ParisTech, 6-8 av Blaise Pascal, Cité Descartes, Champs-sur-Marne, 77455 Marne-la-Vallée, Cedex 2, France.

Later on, this model settled the bases of numerous models, addressing additional features of unsaturated soils behaviour, such as the effects of Lode angle [3], water content [4], anisotropy [5] and degree of saturation [6, 7, 8, 9, 10, 11] among others. Advances on the last point have reintroduced a strong debate dated back to the 60s with Bishop's proposal for an extended effective stress to unsaturated states [12].

This paper presents a thermodynamically consistent framework for hardening plasticity of unsaturated soils. The strain work input to a system with non-connected fluid phases is firstly introduced as an illustration of the separate roles of deformation and drying/wetting processes. The thermodynamics of hardening plasticity of unsaturated soils are then revisited and finally applied to the Barcelona Basic Model [2].

Before addressing the thermodynamics of unsaturated soils, it is worth clarifying the different systems subsequently introduced in this work. The first system is the unsaturated soil itself, being composed of solid particles forming a solid skeleton. The porous volume of this skeleton is filled with two fluid phases (generally a gas and a liquid). These three phases are in contact trough interfaces having their own energy. The second system is obtained by removing from the unsaturated soil the bulk fluids (gas and liquid) whose thermodynamics are separately known. As a consequence, this second system is closed and composed of the solid particles and the interfaces. It will be referred to as the apparent solid skeleton. The last system, which is the one we are interested in, is the previous one from which the interfaces have been removed. It will be called the solid skeleton.

## STRAIN WORK WITH NON-CONNECTED FLUID PHASES

A simplifying assumption is firstly considered in order to illustrate the separate roles of deformation and drying/wetting processes on the contribution to the strain work of the partial porosities changes (see (2)). It is thus assumed that the two fluids fill separate porous networks that are permanently non-connected. This situation could be encountered in soils presenting a well separated double porosity. In this case, if considering also the incompressibility of the solid grains with respect to that of the solid skeleton, that is $d\varepsilon_v = d\phi$, the strain work input to the apparent solid skeleton writes:

$$dw = [p - (1 - s_r) u_a - s_r u_w] d\varepsilon_v + q \, d\varepsilon_q - \phi (u_a - u_w) \, ds_r \qquad (1)$$

where $s_r$ stands fort he Eulerian degree of saturation in water . This relation agrees with the one obtained by Houlsby [13]. According to equation (1), for non-connected fluid phases, the stress couple formed by Bishop's mean stress (with $\chi$ factor identified to the Eulerian water saturation $s_r$), that is $p - (1 - s_r) u_a - s_r u_w$, and the pressure difference $-\phi (u_a - u_w)$ is work conjugate to the strain couple formed by the volumetric strain $\varepsilon_v$ and the Eulerian water saturation $s_r$.

Alternative sets of work conjugate stress and strain can be derived by using equation (1) and the following partition of the Lagrangian porosity:

$$\phi_a + \phi_w = \phi \quad ; \quad \phi_w = s_r \, \phi \quad ; \quad \phi_a = (1 - s_r) \, \phi \qquad (2)$$

together with the condition for solid incompressibility. This end up in different possible choices as suitable stress state variables and may be compared to the work of Fredlund and Morgenstern [14] who showed that any couple of variables among $(p - u_a)$, $(p - u_w)$ and $(u_a - u_w)$ may be used.

## STRAIN WORK WITH CONNECTED FLUID PHASES

When considering connected fluid phases, the analysis of the contribution of each phase to the strain work is not so straightforward. In this case, partial porosities changes $d\phi_a$ and $d\phi_w$ are due not only to changes of the porous volume occupied by each phase but also to the invasion of the volume previously containing one phase by the other phase. To overcome this difficulty, the partial porosities are decomposed into two parts, thus separating invasion and deformational processes:

$$\phi_w = s_r \phi = S_r \phi_0 + \varphi_w \;;\; \phi_a = (1 - s_r) \phi = (1 - S_r) \phi_0 + \varphi_a \;;\; \phi = \phi_0 + \varphi_a + \varphi_w \qquad (3)$$

where $S_r$, $\phi_0$, $\varphi_w$ and $\varphi_a$ respectively stand for the Lagrangian degree of saturation [15], the porosity in the reference configuration and the changes due to deformation only of the Lagrangian porosity for the part of the porous network occupied by water and air. It should be noted that this partition can not be obtained when Eulerian variables are used. In this latter case, changes in $s_r$ may be indifferently due to both drying/wetting and deformation processes. Use of last sub-equation in (3) allows express the work input as:

$$dw = d\omega + d\theta \qquad (4)$$

$$d\omega = p\,d\varepsilon_v + u_a\,d\varphi_a + u_w\,d\varphi_w + q\,d\varepsilon_q \;;\; d\theta = -\phi_0\,(u_a - u_w)\,dS_r \qquad (5)$$

According to (4)-(5), the work input is split into the contribution needed to deform the skeleton, $d\omega$, and that required for the invasion process to occur, $d\theta$. By noting $U$ the fluid-solid interface energy per unit of initial volume, second sub-equation in (5) allows for stating:

$$dU = -\phi_0\,(u_a - u_w)\,dS_r \qquad (6)$$

This relation implies that $U$ and also suction $u_a - u_w$ must be functions of $S_r$ only so that $u_a - u_w = r(S_r)$, which is the classical expression of the water retention curve. This relation does not account for hysteretic effects. Accounting for them is not contradictory to the approach presented here (see e.g. [7]), but requires the consideration of appropriate energy couplings that would weight down the text.

## THERMODYNAMICS OF PLASTICITY

Since the considered system is the apparent solid skeleton, which is a closed system, the Clausius-Duhem inequality which contains the first and the second laws of thermodynamics expresses that, for isothermal evolutions, the strain work input $dw$ to the system has to be greater or equal to the infinitesimal free energy $dF$ that the system can store, the difference $dD$ being spontaneously dissipated into heat. Assuming the solid grains incompressibility and using equations (4) and (5), the Clausius-Duhem inequality writes as follows:

$$dD = (p - u_a)\,d\varphi_a + (p - u_w)\,d\varphi_w + q\,d\varepsilon_q - \phi_0\,(u_a - u_w)\,dS_r - dF \geq 0 \qquad (7)$$

Any further analysis requires the assumptions on the dependency of the free energy $F$ on the state variables. Denoting $\Psi$ the elastic energy, $Z$ the locked energy [16] and $U$ the fluid-solid interfaces energy, the following dependencies are assumed:

$$F = \Psi\,(\varphi_a - \varphi_a^p,\, \varphi_w - \varphi_w^p,\, \varepsilon_q - \varepsilon_q^p,\, S_r) + Z(S_r, \alpha) + U(S_r) \qquad (8)$$

The state equations are obtained considering elastic evolutions thus leading to null plastic strains and an equality in (7). They read:

$$p - u_a = \frac{\partial \Psi}{\partial \varphi_a}; \quad p - u_w = \frac{\partial \Psi}{\partial \varphi_w}; \quad q = \frac{\partial \Psi}{\partial \varepsilon_q}; \quad \phi_0 (u_a - u_w) = -\frac{\partial (\Psi + Z)}{\partial S_r} - \frac{dU}{dS_r} \qquad (9)$$

The first three sub-equations capture to the elastic behaviour of the solid matrix. The last one corresponds to the expression of the water retention curve. It includes the effects of the deformation of the porous volume, except those leading to hysteretic phenomena as stated before. Using state equations (9) and assumptions on dependencies of $F$ given by equation (8), Clausius-Duhem inequality is expressed as:

$$dD = (p - u_a)\, d\varphi_a^p + (p - u_w)\, d\varphi_w^p + q\, d\varepsilon_q^p + \beta\, d\alpha \geq 0 \qquad (10)$$

$$\beta = -\frac{\partial Z(S_r, \alpha)}{\partial \alpha} \qquad (11)$$

$\beta$ is the energy conjugate of the hardening variable $\alpha$, depending on $S_r$. It will be called hardening force and subsequently associated to the limit of elasticity. Introducing the plastic incompressibility of the solid grains which expresses as:

$$d\varphi_a^p = -(1-\chi)\, d\varepsilon_v^p ; \quad d\varphi_w^p = -\chi\, d\varepsilon_v^p \qquad (12)$$

it comes from (10):

$$dD = p^B d\varepsilon_v^p + q\, d\varepsilon_q^p + \beta\, d\alpha \quad \text{where} \quad p^B = p - [1 - \chi(S_r)]\, u_a - \chi(Sr) u_w \qquad (13)$$

where $p^B$ is Bishop's stress. Interestingly, assuming isodeformation of the porous volumes occupied by water and air, that is:

$$d\varphi_w^p / \phi_0 S_r = d\varphi_a^p / \phi_0 (1 - S_r) \qquad (14)$$

and comparing (12) and (14) leads to $\chi(S_r) = S_r$, which is actually the commonly used assumption. According to (13), the Bishop's stress plays the same role in unsaturated states as does Terzaghi's proposal ($p' = p - u$) in saturated conditions. Actually, both definitions arise from solid grains incompressibility assumptions. As a consequence, the current domain of elasticity in triaxial space can be defined by $f(p^B, q, \beta) \leq 0$ where $f$ is the yield criterion, $q$ the deviatoric stress and $\beta$ the hardening force introduced in eq. (10). Assuming associated plasticity for the sake of simplicity, the plastic flow rule is given by:

$$d\varepsilon_v^p = d\lambda \frac{\partial f}{\partial p^B}; \quad d\varepsilon_q^p = d\lambda \frac{\partial f}{\partial q} \qquad (15)$$

As classically done, the plastic multiplier may be obtained invoking the consistency condition ($df = 0$). The framework is completed by giving the plastic modulus:

$$H = -\frac{\partial f}{\partial \beta} \frac{\partial^2 Z}{\partial \alpha^2} \frac{d\alpha}{d\lambda} \qquad (16)$$

The Barcelona Basic Model (BBM) [2] is now revisited in the light of the previous framework to prove its thermodynamical consistency in the line of the demonstration made by Coussy [17] for the modified Cam Clay model. The stress variables used in BBM are the net stress $p - u_a$ and the suction $u_a - u_w$. The key point consists in introducing a capillary hardening in the loading function as:

$$p_0 = p^c \left( p_0^* / p^c \right)^{[\lambda(0)-\kappa]/[\lambda(s)-\kappa]} \tag{17}$$

where $p_0^*$ and $p_0$ are the preconsolidation pressures at saturation and a given suction respectively. It is possible to merge BBM approach into the framework proposed here by a suitable choice of the weighting function $\chi$:

$$\chi(S_r < 1) = 0 \; ; \; \chi(S_r = 1) = 1 \tag{18}$$

The load function of the original Cam-clay model is extended as follows:

$$f = \left( p^B + p_s - 1/2(p_0 + p_s) \right)^2 + q^2/M^2 - 1/4(p_0 + p_s)^2 \tag{19}$$

where $p_s$ is the tensile strength due to suction. Identifying $p_0$ to $\beta$ and $\varepsilon_v^p$ to $-\alpha$, it may be shown that the dissipation finally reads:

$$dD = d\lambda \, (p_0 + p_s)(p_0 - p^B) \tag{20}$$

which remains positive at any time, thus proving the consistency of the model.

# CONCLUSION

A thermodynamically consistent framework for hardening plasticity of unsaturated soils has been presented. In the most general case, three stress variables are needed to entirely describe the evolution of unsaturated soils. However, a reduction in this number may be obtained using constraints on internal deformation. The proposed framework uses a generalized effective stress which includes both the commonly used effective stress with $\chi = S_r$ and the net stress as limiting cases. This last point automatically proves the consistency of the Barcelona Basic Model as well as that of all the models derived from it by taking a smooth function of $S_r$ for the weighting function $\chi$.